\documentclass[aps,prl,twocolumn,preprintnumbers]{revtex4}
\usepackage{graphicx,amsmath}
\usepackage{dcolumn}
\usepackage{amsmath}
\usepackage{amssymb}
\usepackage{amsfonts}

\begin{document}

\title{Thermodynamics in Terms of a
Sequence of $n-$chains Derived from a Martingale Decomposition
of the Energy Process}         
\author{David Ford}        
\email[]{dkf0rd@netscape.net}
\affiliation{Department of Physics, Naval Post Graduate School, Monterey, California}
\date{\today}          
\begin{abstract}
The role of the algebraic method has long been understood in shedding light
on the topological structure of sets. However, when the set is a simplicial complex
and host to a dynamical process,
in particular the trajectory of a canonically distributed system in thermal equilibrium
with a heat bath, the algebra re-enters. Via a theorem of L$\acute{\mbox{e}}$vy and Dynkin,
there is a correspondence between a system's energy process at equilibrium
and a sequence of $n-$chains on the state space.

\end{abstract}
\maketitle

\section{Fundamentals and Review}

\subsection{Thermostatics}

Let  the set  $\{\omega_1, \omega_2, \ldots, \omega_N \}$
be the state space for a random process $\{ x_t: t \in [ 0, \infty) \}.$
According to the results of \cite{application}, under the assumption that
the ``position'' process $x_t$ is a trajectory of a canonically distributed
subsystem in equilibrium with a heat bath at temperature
$\theta$, there is a well defined energy process

\begin{equation}\label{h.} 
h(x_t)=\theta \, \log \Big( \,\frac{\Pi}{ p(x_{t})} \,\Big)
\end{equation}

\noindent where,

\begin{equation}\label{Pi}
\Pi = \Big( p_{1}p_{2} \cdots p_{N} \Big) ^{\frac{1}{N}}
\end{equation}
\noindent and for $k=1, 2, \ldots, N$
$$
p_k =  \, p(x_t  \in \omega_k)~.
$$

Further \cite{surface}, if the $N-$state process $x_t$ has a characteristic Carlson depth
(in the finite state Markoff case this is simply the $L^1$ norm of the cycle between
visits to a rare state) we have 
 
\begin{figure}[htbp]
\begin{center}
\leavevmode
\includegraphics[width=60mm,keepaspectratio]{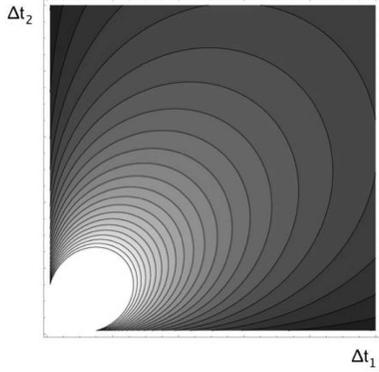}
\caption{Surfaces of Constant Temperature for a Two State System}
\label{greyisotemps}
\end{center}
\end{figure}

 \begin{widetext}
\begin{equation}\label{daformula}
\theta(\Delta \mathbf{t})= \frac{ \textrm{const.} }
{ \| t \|_2 \,  \sqrt{ (\log [ \frac{\breve{\Pi} }{\Delta t_1} ])^2+(\log [ \frac{ \breve{\Pi} }{\Delta t_2} ])^2+\cdots+
(\log [ \frac{\breve{\Pi} }{\Delta t_{N-1}} ])^2+1        }}
\end{equation}
\end{widetext}

\noindent where,
$\Delta t_k$
is the characteristic time associated with the $k^{th}$
state during a cycle and 
$$
\breve{\Pi} = \Big(\Delta t_{1}\Delta t_{2} \cdots \Delta t_{N} \Big) ^{\frac{1}{N}}~.
$$
\noindent For a simple 
example see \cite{glauber}. Sample Isotherms as seen
from the time domain are shown in figure
\ref{greyisotemps}.

With  these microscopic quantities in hand, one is in a position to
calculate the free energy and many other important
quantities. The internal energy $$U = \mathbf{p} \cdot  \mathbf{h} $$
\noindent will be of particular importance in the next 
subsection.

\subsection{Thermodynamics}

It may happen that  there are changes to the observed probabilities, changes
in the length of the characteristic time scale or both. if it is assumed that
these changes are occurring in a subsystem that is in a state of continual
equilibrium with a heat bath (or perhaps the observer shifts her attention from one
subsystem to another all in established equilibrium with the same bath), the notions
of work done, heat gained or lost by the system become applicable. 

In equilibrium these 
changes are given by the product rule for differentiation
(the {\bf{First Law}}) as

\begin{equation}\label{rate}
 \dot{U} =  \underbrace{ \mathbf{p} \cdot \dot{ \mathbf{h}} }_{work \, rate}
+ \underbrace{ \dot{ \mathbf{p}} \cdot  \mathbf{h} }_{heat \, rate}
\end{equation}

In the nonstationary case equation (\ref{Pi}) transforms in the obvious
way from $\Pi$ into $\Pi(t)$. Similarly $h(x_t)$ becomes $h(x_t, t)$, etc.
The calculation of the rates involve time derivatives of these
quantities and 
are easily computed. The vectors appearing
in the dot products on the RHS of equation
(\ref{rate}) are typically high dimensional.

Fortunately, in the statistical mechanical context, the  work rate admits a low dimensional
description as well. Typically this is system dependent. Famous examples
include
pressure and volume changes or magnetization and changes in an applied field.
The low-dimensional description of the heat rate is often system
independent and is given in terms of temperature and entropy.

Frequently, as  the microscopic dynamics often occur on state spaces with a topology
these same macroscopic phenomena may also admit a second, low-dimensional 
interpretation in terms of their interplay with features of the state space.
The purpose of this paper is to illustrate the twofold nature of the thermodynamics.
The connection is accomplished via the subsystem's trajectory through its path
space and the path decomposition of L$\acute{\mbox{e}}$vy and Dynkin.

\subsection{Dynkin-L$\acute{\mbox{e}}$vy Decomposition of the Path} 
This subsection focuses some elementary aspects of the martingale
theory on applications to the energy process. In particular, the
context is that  of a  canonically distributed subsystem's
trajectory through its state space.
For readers unfamiliar with the martingale point of view an 
excellent reference for this section is \cite{joe}.

Recall that according to the L$\acute{\mbox{e}}$vy Formula, see for example \cite{bremaud}, the 
martingale decompositon \cite{joe} of a Markoff process
is given by:

$$
g(x_{t}) - g(x_{0}) - \int_{0}^{t} \sum_{k \neq x_{s}} q_{\,x_{s} \, k} \left( g(k)-g(x_{s}) \right) ds     
$$

\noindent here $$q_{\,x_{s} \, k} = \lim_{s^+   \rightarrow s} \frac{P(x_{s^+} = k \mid x_s )}{ s^+ - s}$$
is the generator of the process.

Of particular interest in the sequel  is the case when $g(\cdot)$ is the energy function
$h(x_t, t)$ and the temperature is held constant. Under these conditions the Dynkin-L$\acute{\mbox{e}}$vy kernel
becomes 
$$\theta \, \sum_{k \neq x_{s}} q_{_{x_{_{s}}  k}}\mbox{\scriptsize{($s$)} }
 \left( \log\frac{\Pi(s^+)}{p(k, s^+)} - \log\frac{\Pi(s)}{p(x_s, s)} \right).
$$
Note: it may also be that the generator is itself a function of time. 

\subsection{On the Choice of State Space}

As the main purpose of this paper is to bring out 
the connection between the physical and 
and algebraic descriptions of the subsystem's trajectory
 it is worthwhile to choose a state space
that is meaningful in either context. 

Exchange rules
in the context of urn models, \cite{siegert}, \cite{GandL}, \cite{LandD}, are
frequently used as state spaces to study diffusion,
spin flip dynamics, granular materials, phase transitions, etc.
Further the resulting state spaces
are often simplexes or simplicial complexes 
which have long been used as building blocks in the study
of algebraic geometry and topology.

\begin{figure}[htbp]
\begin{center}
\leavevmode
\includegraphics[width=60mm,keepaspectratio]{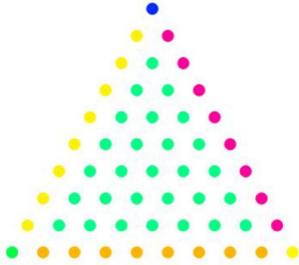}
\caption{A basic example of a $2-$simplex derived from the state space for Gordon-Newell
exchange.  The $0-$chain, $1-$chain and $2-$chain shown (color coded)
are derived from the  martingale decomposition of the local energy process.}
\label{simplex1}
\end{center}
\end{figure}

A particularly simple member of this family of exchange models
is the Gordon-Newell queue \cite{gordon-newell}.
For an interpretation of that process in terms of its thermodynamics
see \cite{application}.  Figure \ref{simplex1} illustrates the algebraic
topology associated with its dynamics. Despite the potential
for an arbitrarily large number
of non-degenerate (energy levels are
distinct)  microstates,
both the thermodynamic and
algebraic descriptions of the energy process are three dimensional.
The connection will be more fully
developed in the sequel. 

\subsection{Elements of the Gordon-Newell Process}

The state space of a Gordon-Newell process may be represented
as a matrix. The dimension of the column space is the number
of urns. The rows represent the states accessible to the system
and may be used as state labels.

For an instantiation based on $M $ balls and $m $ urns, the cardinality
of the state space is simply the number of ways to put 
$M $ balls (indistinguishable) in $m $ urns. In equation (\ref{gnmatrix}) an example of the state space matrix and
row labeling scheme are presented. 
The left hand
side of equation (\ref{gnmatrix}) defines the list of state labels associated
with a space based on three urns and three balls (a total
of ten states). The right hand side of the equation lists the states
themselves. The dimension of the column space is equal to the number of urns.

In the sequel there will be essentially two types of vector dot products, those
describing macrostructure and those describing microstructure.
For clarity, dot products with respect to the low dimensional row
space will be denoted with a circled circle notation, for example
(see equation (\ref{gnmatrix}))
$$\pmb{\alpha}(7) \circledcirc \pmb{\eta} = 1\eta_{1} +  2 \eta_{2} + 0 \eta_{3}. $$

\noindent Dot products in the  column space and other high dimensional vectors
will use the standard `` $\pmb{\cdot}$ '' notation.  

 The labels $\pmb{ \alpha}(1), \pmb{ \alpha}(4)$
and $\pmb{ \alpha}(10)$ are the $0-$simplexes. These are the states with
all the balls in one urn or, alternatively,  the states with two urns empty. The states
$\pmb{ \alpha}(2)$ and $\pmb{ \alpha}(3)$ belong to the same $1-$simplex, namely
the set of states with the first urn empty and the others non-empty.
The label $\pmb{ \alpha}(6)$  state belongs to the interior of the region. That is,
the set of states with no empty urns. 

\begin{equation}\label{gnmatrix} 
\;
\begin{array}{ cc}
\pmb{ \alpha}(1) &\\
 \pmb{ \alpha}(2) &\\
 \pmb{ \alpha}(3)  &\\
 \pmb{ \alpha}(4) &\\
 \pmb{ \alpha}(5)  &\\
 \pmb{ \alpha}(6)   &\\
 \pmb{ \alpha}(7)  &\\
 \pmb{ \alpha}(8)  &\\
 \pmb{ \alpha}(9)  &\\
 \pmb{ \alpha}(10) & 
\end{array}
  \,
=
\; \;\;
\underbrace{
  \begin{array}{ c c c}
 0 & 0 & 3 \\
 0 & 1 & 2 \\
 0 & 2 & 1 \\
 0 & 3 & 0 \\
 1 & 0 & 2 \\
 1 & 1 & 1 \\
 1 & 2 & 0 \\
 2 & 0 & 1 \\
 2 & 1 & 0 \\
 3 & 0 & 0 
  \end{array} 
  }_{\mbox{3 urns}}
\end{equation}

A basic version of the exchange rule consists of an $m \times m$ global routing table
(a transition probability matrix
for an embedded discrete process)
and $m-$timescales that characterize (globally) the rate of activity of the individual urns. 

Let $P_{ij}$ denote the probability that the discrete process transitions a single ball
from urn ``i'' to urn ``j''.
The stationary vector (eigenvalue $1$) of the transition matrix will be denoted
$\{ \pi_1, \pi_2, \ldots, \pi_m \}.$ The urn rates will be denoted
$\{ q_1, q_2, \ldots, q_m \}.$ 

Any component of a stationary measure of the process
may be given in terms of these parameters:
$$ ( \frac{\pi_{1}}{q_{1}})^{\alpha_{1}} (\frac{\pi_{2}}{q_{2}})^{\alpha_{2}} 
\cdots (\frac{\pi_{m}}{q_{m}})^{\alpha_{m}}
$$
\noindent  where, $\{ \alpha_{1}, \alpha_{2}, \ldots,\alpha_{m} \}$ are entries
in that row of the state space matrix associated with the state label 
$\pmb{\alpha}(\cdot)$.

A slightly more flexible version of the exchange rule allows the routing table
and rates to vary locally from state to state. In addition, the parameters may be functions
of time.

\section{The Connection with the Heat Rate Process}

\subsection{The Constant Temperature, Time Stationary Equilibrium}
With the state energies and temperature given in terms of the
canonically distributed subsystem's statistics and timescales by
equations (\ref{h.}) and (\ref{daformula}), a stationary measure $\pmb{\mu}(\cdot)$ of the 
exchange process may be expressed (componentwise) in a few ways

\begin{equation}\label{mu1} 
\mu(\pmb{\alpha}) =
\left\{ \begin{array}{ll}
( \frac{\pi_{1}}{q_{1}})^{\alpha_{1}} (\frac{\pi_{2}}{q_{2}})^{\alpha_{2}} 
\cdots (\frac{\pi_{m}}{q_{m}})^{\alpha_{m}}&
\\
\\
e^{- (\eta_1 \alpha_{1} +\eta_2 \alpha_{2}+\ldots+\eta_m \alpha_{m} )} &
\\
\\
e^{- \frac{h(\pmb{\alpha})}{\theta}}. &
\end{array} \right.  
\end{equation}
\noindent Recall that the top expression on the RHS of equation (\ref{mu1})
is the stationary measure in terms of the exchange rule parameters
and was introduced in the last section. The middle expression follows 
using the substitution 
$$
\eta_. = - \log( \frac{\pi_{.}}{q_{.}})~.
$$ 
\noindent The bottom expression gives a stationary measure
in terms of energy and temperature. 

As is apparent from equation (\ref{daformula}), the temperature
may vary (via dilatation of the time scales) even though the state probabilities
themselves are constant in time. When both the temperature and probabilities
are constant in time
\begin{equation}\label{h.2}
h(x_t, t ) = 
\left\{ \begin{array}{ll}
h(x_t) &
\\
\\
\theta  \, \pmb{\eta} \circledcirc ( \pmb{\alpha}(x_t) - \bar{\pmb{\alpha}} )&
\end{array} \right.  
\end{equation}

\noindent where $\bar{\pmb{\alpha}}$ is the arithmetic mean of the rows of the state space matrix.
For the example given in equation (\ref{gnmatrix}) this vector is $\{ 1, 1, 1\}.$

Dynkin-L$\acute{\mbox{e}}$vy
kernel for the energy process is given by

\begin{equation}\label{DLorig} 
\sum_{k \neq x_{s}} q_{\,x_{s} \, k} \; \theta
\left\{ \begin{array}{ll}
\pmb{\eta} \circledcirc ( \pmb{\alpha}(x_s) - \pmb{\alpha}(k) ) &
\\
\\
\frac{h(\pmb{\alpha}(x_s) ) - h(\pmb{\alpha}(k) ) }{\theta} &
\end{array} \right.  
\end{equation}
\noindent In what follows the Dynkin-L$\acute{\mbox{e}}$vy coefficient
at a site ``$\ast$'' will be denoted $dl(\ast)$.

\subsection{Physical Significance of the Dynkin-L$\acute{\mbox{e}}$vy Kernel for the
Energy Process }

Under the hypothesis of this subsection, constant temperature, time stationary
equilibrium,
the Dynkin-L$\acute{\mbox{e}}$vy kernel takes the form given in equation (\ref{DLorig}).
It may be shown that this corresponds to the heat process associated with the
energy process $h(x_t).$ For

\begin{eqnarray}
\langle dl(x_s) \rangle ds &=&\sum_{x_s} \sum_k \,
        p(x_s, s) \, P_{x_s \, k}(s, s^+)\nonumber \\
 & &\times \left[  \pmb{\alpha}(k) - \pmb{\alpha}(x_s) \right]
       \circledcirc \pmb{\eta}\nonumber \\
 &=& \sum_k p(k, s+) \left[  \pmb{\alpha}(k) - \bar{\pmb{\alpha}}
            \right]  \circledcirc \pmb{\eta}\nonumber \\
  & & -\sum_{x_s} p(x_s, s) \left[  \pmb{\alpha}(x_s) - \bar{\pmb{\alpha}}
            \right]  \circledcirc \pmb{\eta} \nonumber \\
 &=& \sum_m \left( p(m, s+) - p(m, s) \right)\nonumber \\
 & & \times \left[  \pmb{\alpha}(m) - \bar{\pmb{\alpha}}
            \right]  \circledcirc \pmb{\eta}~.\nonumber
\end{eqnarray}
So that,
\begin{eqnarray}
\langle dl(x_s) \rangle &=&
 \sum_m \dot{p}(m, s ) h(m, s)~.      
\end{eqnarray}
\noindent This is of course the heat rate's contribution to the ``First Law'',
equation (\ref{rate}).

\subsection{Algebraic Significance of the Dynkin-L$\acute{\mbox{e}}$vy Kernel for the
Energy Process }

Inspection of equation (\ref{DLorig}) shows that the Dynkin-L$\acute{\mbox{e}}$vy coefficient at a site
depends on the $\pmb{\alpha}$ differences with the states to which the process may
transition, weighted by the probability rate that the transition takes place during a small
time $s \rightarrow s^+$.

Recall from the previous section that in
 the most basic version of the exchange rule there is one global
matrix (site independent) that governs the transitions of the embedded
discrete process and one set of  inverse characteristic times that
governs the rates of the urns. 

\begin{figure}[htbp]
\begin{center}
\leavevmode
\includegraphics[width=60mm,keepaspectratio]{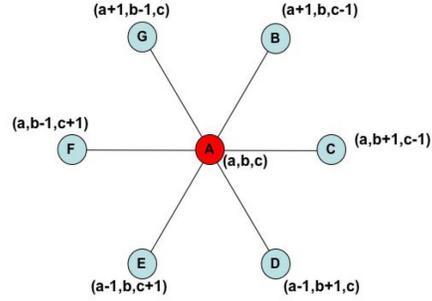}
\caption{Typical Neighborhood of an Interior Point in the State Space}
\label{simplexnbd}
\end{center}
\end{figure}

A typical local neighborhood of an interior point ``A'' with state
label $\pmb{\alpha}(A)= \{a, b, c\}$ is shown in figure \ref{simplexnbd}.
The quantities relevant to the calculation of the Dynkin-L$\acute{\mbox{e}}$vy coefficient
at site ``A'' are listed in the table. 

\begin{table}[htbp]
\caption{A break down of the summands that comprise a Dynkin-L$\acute{\mbox{e}}$vy coefficient:
possible transition states, $\alpha-$index differences and corresponding 
small time probabilities of transition.
See figure \ref{simplexnbd}.}
\hspace{45mm}
\begin{tabular}{lcc}
$k$&
$\;\;\;\;\;\; \alpha(k) - \alpha(A)$ \hspace{7mm} &
$\mbox{ \Large p}_{A \, k}\mbox{\scriptsize{(s,$ s^+$)} }$\\
\vspace{-3mm}\\
\hline
A  & $ 0, 0, 0 $ & 1 - ( \;\;\; ) \\
B & $ 1, 0, -1$& $q_3 P_{31}$ ds\\
C  & $ 0, 1, -1$& $q_3 P_{32}$ ds\\
D  & $ -1, 1, 0$& $q_1 P_{12}$ ds\\
E  & $ -1, 0, 1$&  $q_1 P_{13}$ ds\\
F  & $ 0, -1, 1$&  $q_2 P_{23}$ ds\\
G  & $ 1, -1, 0$&  $q_2 P_{21}$ ds\\ 
\hline 
\end{tabular}
\end{table}

Notice that the table entries are 
invariant with respect to the actual location of site ``A''. That is,
the Dynkin-L$\acute{\mbox{e}}$vy coeficients are constant on the interior of the simplex.
A similar situation holds in the case ``A'' belongs to one of the boundaries. 

Consider the
$1-$simplex of states that have the first urn empty and the other two urns
non-empty (for a picture see the orange 1-boundary in figure \ref{simplex1}).
The typical neighborhood and table for this boundary are obtained from
those shown above for the interior points by simply deleting points ``D'' and ``E''.
The same table and figure result for all points on this face. In
this way a real valued coefficient is associated with each $1-$simplex
in the complex.
The set of these coefficients specifies a particular $1-$chain.

The reader should be able to convince herself that under
the conditions of the basic exchange rule: one global (site independent)
transition matrix and one global set of characteristic times for the urns,
the Dynkin-L$\acute{\mbox{e}}$vy coefficients for the energy process specify $n-$chains
in higher dimensions as well.

\section{The Connection with the Work Rate Process}

It may happen, see equation(\ref{daformula})
or figure \ref{greyisotemps},  that the probabilities vary in time but the system
remains on a constant temperature surface. In this situation
the quantities $\pi_{\pmb{.}}, q_{\pmb{.}}$ and $\eta_{\pmb{.}}$
from equation (\ref{mu1})
become functions of time.  The Dynkin-L$\acute{\mbox{e}}$vy kernel of the
energy process takes the form
(see the appendix, equation (\ref{appDL}) )

\begin{eqnarray}\label{DLnext} 
dl(x_s)&=& - \sum_{k \neq x_{s}} q_{\,x_{s} \, k}(s)\nonumber \\ 
& & \times \left( h(\pmb{\alpha}(x_s),  s) - h(\pmb{\alpha}(k),  s^+) \right)  \nonumber \\
 \nonumber \\
&=& \theta \; \dot{\pmb{\eta} }(s) \circledcirc \left( \pmb{\alpha}(x_s) - 
\pmb{\bar{\alpha}} \right)\nonumber \\
& & + \theta \, \sum_{k \neq x_{s}} q_{\,x_{s} \, k}(s)\nonumber \\
& & \times  \pmb{\eta}(s) \circledcirc ( \pmb{\alpha}(k) - \pmb{\alpha}(x_s) )\nonumber \\
\end{eqnarray}

Comparison of (\ref{DLnext}) with equation (\ref{DLorig}) reveals an additional term
$$ \theta \, \dot{\pmb{\eta} }(s) \circledcirc \left( \pmb{\alpha}(x_s) - 
\pmb{\bar{\alpha}} \right)~.$$

\noindent This site dependent term is clearly the local work rate at $x_s$. For
(see equation (\ref{rate}))
\begin{eqnarray}\label{workrate} 
 \theta & & \hspace{-7mm} \sum_{x_s \in \Omega} p(x_s) \left[
 \dot{\pmb{\eta} }(s) \circledcirc \left( \pmb{\alpha}(x_s) - 
\pmb{\bar{\alpha}} \right) \right] \nonumber \\
&= &  \theta \,\sum_{m} p(m) \left[
\dot{\pmb{\eta} }(s) \circledcirc \left( \pmb{\alpha}(m) - 
\pmb{\bar{\alpha}} \right) \right] \nonumber \\
&=& \pmb{p} \cdot \pmb{\dot{h}}~.
\end{eqnarray}

\section{summary}
Under the condition of global (site independent) exchange rules on a 
simplicial complex, a correspondence
has been shown between the chain groups (with real coefficients) of the state space
and the thermodynamics. In the case of constant temperature, time stationary
equilibrium, the Dynkin-L$\acute{\mbox{e}}$vy kernel ties the system's local heat rate to specific members
of those groups.

Under the same conditions, in the case of constant temperature but time varying dynamics,
 it is shown
that, due to the presence of a site dependent  (local) work rate term, the correspondence with the
chain groups is lost. Thus, the connection has been made between transitions from one thermal
equilibrium to another and from one sequence of p-chains on the state space to another.

\appendix
\section{Appendix A: \, Derivation of the Dynkin-L$\acute{\mbox{E}}$vy Kernel for the Energy Process}

Let $h(x_t, t)$ denote the energy of the state $x_t$ at the time $t$. That is,
the Hamiltonian for the system may have an explicit time dependence.
The collection of all system trajectories (the position processes $\{x_t, 
t \in [ 0, \infty ) \} $) will be denoted $\Omega$. Let ${\cal{P}}$ denote
the probability on path space the describes the trajectories of the position process.

The energy difference between the an ``initial'' state $x_{t_0}$ at time
$t_0$ and a ``final'' state $x_{t_m}$ at time $t_m$, may be decomposed into a
telescoping sum:
\begin{eqnarray}
\delta h\mid_{t_0}^{ t_m}&=& h(x_{t_m}, t_m) - h(x_{t_0}, t_0) \nonumber \\
&=&\left[ h(x_{t_m}, t_m) - h(x_{t_{m-1}}, t_{m-1}) \right]  \nonumber \\
& &+\left[h(x_{t_{m-1}}, t_{m-1}) - h(x_{t_{m-2}}, t_{m-2}) \right]\nonumber \\
& &+ \ldots \nonumber \\
& &+ \left[ h(x_{t_1}, t_1) - h(x_{t_0}, t_0) \right]~. 
\end{eqnarray}
As is customary, let ${\cal{F}}_t$ be the  $\sigma-$algebra generated by
paths in $\Omega$ up to time $t$. For simplicity, is supposed that ${\cal{F}}_0$ 
contains the information about the explicit time dependence of the Hamiltonian.

Also, as is customary, let $E( \ast \vert {\cal{F}}_t ) $ denote the projection operator
(conditional expectation) that projects its argument onto the set of square integrable
functions on the measurable space $(\Omega, {\cal{F}}_t  )$ using the measure
${\cal{P}}$. 
The accumulated difference between the actual trajectory and its projections may
be approximated as follows

\begin{eqnarray}
\perp h\mid_{t_0}^{ t_m}&=& \left[ h(x_{t_m}, t_m) - h(x_{t_{_{m-1}}}, t_{_{m-1}}) \right] \nonumber \\
&-& E(  h(x_{t_m}, t_m)\nonumber \\
& & \;\;\;\; - h(x_{t_{_{m-1}}}, t_{_{m-1}})    \vert  \, {\cal{F}}_{t_{_{m-1}}} )\nonumber \\
&+&\left[h(x_{t_{_{m-1}}}, t_{_{m-1}}) - h(x_{t_{_{m-2}}}, t_{_{m-2}}) \right]\nonumber \\
&-& E( h(x_{t_{_{m-1}}}, t_{_{m-1}}) \nonumber \\
& & \;\;\;\;- h(x_{t_{_{m-2}}}, t_{_{m-2}})    \vert  \, {\cal{F}}_{t_{_{m-2}}} )\nonumber \\
&+& \; \cdots \nonumber \\
&+& \left[ h(x_{t_1}, t_1) - h(x_{t_0}, t_0) \right] \nonumber \\
&-& E( h(x_{t_{1}}, t_{1}) - h(x_{t_{0}}, t_{0})    \vert  \, {\cal{F}}_{t_{0}} )
\end{eqnarray}

It is a standard result that $\perp h\mid_{t_0}^{ t_m}$ is a martingale with 
respect to the filtration $ {\cal{F}}_{t} $ previously described.

In the case that the position process is Markoff, the small time contributions
to the kernel may be computed as

\begin{eqnarray}
E( && \hspace{-6mm}h(x_{t_{m + 1}}, t_{m+1}) - h(x_{t_{m}}, t_{m})    \vert  {\cal{F}}_{t_{m}} ) \nonumber \\
\nonumber \\
&=&  [ h(x_{t_{m}}, t_{m+1}) - h( x_{t_{m}},  t_{m} ) ]\nonumber \\
& & \times(1 - O(t_{m+1} - t_{m})) \nonumber \\
& & + \sum_{k \neq x_{t_m}} \left[   h(k, t_{m+1}) - h( x_{t_{m}} , t_{m})\right]\nonumber \\
& & \times \mbox{ \Large P}_{ \hspace{-2mm} \,x_{t_m} \, k} 
\mbox{ \hspace{-2mm}  \scriptsize{($t_{m-1}$, $t_m$)} }
\end{eqnarray}

more precisely:

\begin{eqnarray}
E( && \hspace{-6mm}h(x_{t_{m + 1}}, t_{m+1}) - h(x_{t_{m}}, t_{m})    \vert  {\cal{F}}_{t_{m}} ) \nonumber \\
\nonumber \\
&=& \frac{ h(x_{t_{m}}, t_{m+1}) - h( x_{t_{m}},  t_{m} ) }{ t_{m+1} - t_{m} } (t_{m+1} - t_{m}) \nonumber \\
& &+ \sum_{k \neq x_{t_m}} \left[   h(k, t_{m+1}) - h( x_{t_{m}} , t_{m})\right] \nonumber \\
& & \times \mbox{ \Large q}_{  \,x_{t_m} \, k} 
\mbox{ \hspace{-2mm}  \scriptsize{( $t_m$)} }
\mbox{ ($t_{m+1} - t_{m}$) }~.
\end{eqnarray}

Assuming that
$$ h(k, s^+) - h( k ,s)$$ 
is small. for example h as a differentiable function of time, this is 

\begin{eqnarray}
E( && \hspace{-6mm}h(x_{t_{m + 1}}, t_{m+1}) - h(x_{t_{m}}, t_{m})    \vert  {\cal{F}}_{t_{m}} ) \nonumber \\
\nonumber \\
&=&   
\frac{ h(x_{t_{m}}, t_{m+1}) - h( x_{t_{m}},  t_{m} ) }{ t_{m+1} - t_{m} } (t_{m+1} - t_{m}) \nonumber \\
& &+ \sum_{k} \left[   h(k, t_{m}) - h( x_{t_{m}} , t_{m}) \right]\nonumber \\
& & \times \mbox{ \Large q}_{  \,x_{t_m} \, k} 
\mbox{ \hspace{-2mm}  \scriptsize{( $t_m$)} }
\mbox{ ($t_{m+1} - t_{m}$) }
\end{eqnarray}

Under these conditions the accumulated affects over the the time interval
$[0,t]$ are

\begin{eqnarray}
\int_0^t  \bigg(  \frac{ \partial h(x_{s}, s)  }{\partial s}
&+& \sum_{k} \left[   h(k, s) - h( x_{s} , s) \right]\nonumber \\
& & \times \mbox{ \Large q}_{  \,x_{s} \, k} 
\mbox{ \hspace{-2mm}  \scriptsize{($s$)} } \bigg)
\; ds
\end{eqnarray}

At constant $\theta$, in terms of the exponents from equation (\ref{h.2})
\begin{eqnarray}\label{appDL}
\theta & & \hspace{-7mm}
\int_0^t  \dot{\eta}(s) \circledcirc [ \alpha(x_s) - \bar \alpha ]\nonumber \\
&+& \sum_{k} \eta(s) \circledcirc \left[ \alpha(k) - \alpha(x_s) \right]\nonumber \\
& & \times \mbox{ \Large q}_{_{\,x_{s} \, k}} 
\hspace{-1.5mm}  \mbox{ \scriptsize{(s)} }
\; ds\end{eqnarray}

At constant $\theta$, in terms of fancier exponents resulting from the application
of time varying microforces (local exchange rules and timescales)
\begin{eqnarray}
\theta \int_0^t   ds
& & \hspace{-6mm} \bigg(
\dot{\eta}(s) \circledcirc [ \alpha(x_s) - \bar \alpha ]\nonumber \\
&+& \dot{\delta}(x_s, s) \circledcirc \alpha(x_s) - 
\overline{ \dot{\delta}(\ast, s) \circledcirc \alpha(\ast)} \nonumber \\
&+& \sum_{k} \mbox{ \Large q}_{_{_{x_s k}} }
\hspace{-1.5mm}\mbox{\tiny{(s)}  }
\; \bigg[
 \eta(s) \circledcirc \left[ \alpha(k) - \alpha(x_s) \right] \nonumber \\
& &+  \delta(k,s) \circledcirc \alpha(k) -  \delta(x_s,s) \circledcirc \alpha(x_s) 
\bigg]
\bigg) \nonumber \\
\end{eqnarray}

\end{document}